\documentclass[reprint, prl, aps, twocolumn, floatfix, showkeys]{revtex4-1}

\usepackage{amssymb}
\usepackage{amsmath}
\usepackage{graphicx}
\usepackage{bm}
\usepackage{dcolumn}
\usepackage{color}
\usepackage{ulem}
\usepackage{float}
\usepackage{textcomp}
\restylefloat{table}
\usepackage{verbatim}
\usepackage{epstopdf}
\usepackage[colorlinks]{hyperref}
\usepackage[mediumspace,mediumqspace,squaren]{SIunits}

\begin{document}

\title{Observation of Spontaneous Ferromagnetism in a Two-Dimensional Electron System}
\date{\today}

\author{Md.\ S. Hossain}
\author{M. K.\ Ma}
\author{K.\ A. Villegas-Rosales}
\author{Y. J.\ Chung}
\author{L. N.\ Pfeiffer} 
\author{K. W.\ West}
\author{K. W.\ Baldwin}
\author{M.\ Shayegan}
\affiliation{Department of Electrical Engineering, Princeton University, Princeton, New Jersey 08544, USA}

%%%%%%%%%%%%%%%%% END OF PREAMBLE %%%%%%%%%%%%%%%%

\begin{abstract}

What are the ground states of an interacting, low-density electron system? In the absence of disorder, it has long been expected that as the electron density is lowered, the exchange energy gained by aligning the electron spins should exceed the enhancement in the kinetic (Fermi) energy, leading to a (Bloch) ferromagnetic transition. At even lower densities, another transition to a (Wigner) solid, an ordered array of electrons, should occur. Experimental access to these regimes, however, has been limited because of the absence of a material platform that supports an electron system with very high-quality (low disorder) and low density simultaneously. Here we explore the ground states of interacting electrons in an exceptionally-clean, two-dimensional electron system confined to a modulation-doped AlAs quantum well. The large electron effective mass in this system allows us to reach very large values of the interaction parameter $r_s$, defined as the ratio of the Coulomb to Fermi energies. As we lower the electron density via gate bias, we find a sequence of phases, qualitatively consistent with the above scenario: a paramagnetic phase at large densities, a spontaneous transition to a ferromagnetic state when $r_s$ surpasses 35, and then a phase with strongly non-linear current-voltage characteristics, suggestive of a pinned Wigner solid, when $r_s$ exceeds $\simeq 38$. However, our sample makes a transition to an insulating state at $r_s\simeq 27$, preceding the onset of the spontaneous ferromagnetism, implying that, besides interaction, the role of disorder must also be taken into account in understanding the different phases of a realistic dilute electron system. 

\end{abstract}

\maketitle

The ground state of an interacting, dilute electron system and its magnetic properties have long been topics of great interest in many-body physics \cite{Bloch.1929, Wigner.1934, Stoner.1947, AshcroftMermin, Tanatar.1989, Attaccalite.PRL.2002, Drummond.PRL.2009}. At low densities the interaction energy dominates over the kinetic energy. A disorder-free, itinerant electron system at sufficiently low electron densities is expected to make a transition to a ferromagnetic ground state \cite{Bloch.1929}, namely a Bloch ferromagnet. At even lower densities, electrons are predicted to condense into a Wigner solid, as ordered array of electrons \cite{Wigner.1934}. For a quantitative description, it is convenient to characterize the interacting electron system with the dimensionless parameter $r_s$, the average inter-electron distance in units of the effective Bohr radius (equivalently, $r_s$ is also the ratio of the Coulomb to Fermi energies). In a three-dimensional metal, $r_s$ values are typically well below those required for the ferromagnetic transition \cite{AshcroftMermin}. For a two-dimensional electron system (2DES), Monte Carlo calculations \cite{Attaccalite.PRL.2002} predict a transition to full magnetization when $r_s$ exceeds $26$, followed by another transition to a Wigner crystal state for $r_s > 35$. (These predictions are not all corroborated by other Monte Carlo calculations; see, e.g., Ref. \cite{ Drummond.PRL.2009}.) Achieving a very dilute and clean 2DES so that interaction phenomena are not hindered by disorder and single-particle localization, however, is extremely challenging. For a 2DES, we have $r_s = (m e^2/4 \pi \hbar^2 \varepsilon \varepsilon_0)/ (\pi n)^{1/2}$, where $m$ is the electron ``band" effective mass (in units of the free electron mass), $\varepsilon$ is the dielectric constant, and $n$ is the 2DES density. For example, in a GaAs 2DES ($m=0.067$ and $\varepsilon=13$), $r_s\simeq26$ corresponds to a density of $n=4.6 \times 10^8$ cm$^{-2}$, which is indeed very difficult to attain \cite{Sajoto.PRB.1990, Zhu.PRL.2003}. In GaAs 2D \textit{hole} systems ($m \simeq 0.4$), large $r_s$ values can be reached more easily, and in fact hints of a Wigner solid formation near $r_s \simeq 35$ were reported \cite{Yoon.PRL.1999}. However, partly because of the strong spin-orbit interaction, the extraction of spin polarization of 2D holes is not straightforward \cite{Papadakis.Science.1999, Tutuc.PRL.2001, Winkler.PRB.2005}.

\begin{figure*}[t!]
\includegraphics[width=1\textwidth]{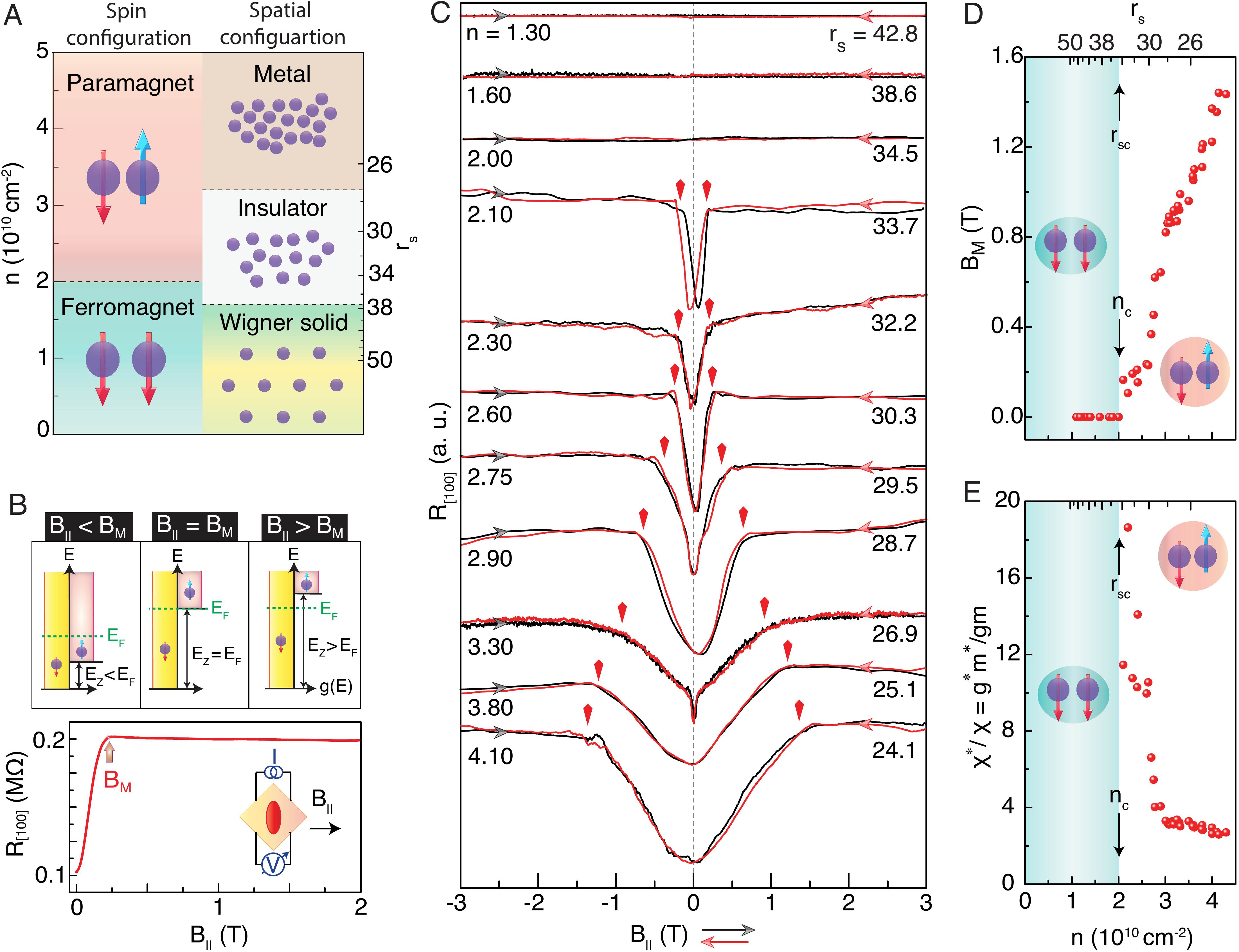}
\caption{\label{fig:fig1} Highlights of our experimental findings. (\textit{A}) An experimental phase diagram, suggested by our study, for the ground states of an interacting 2DES at zero magnetic field as it is made progressively more dilute. The sample is paramagnetic at high densities, and its resistance vs. temperature dependence switches from metallic to insulating below a density of $3.2$ (in units of $10^{10}$ cm$^{-2}$). As the density is further lowered, the 2DES makes an abrupt transition to a fully-magnetized state below $2.00$; this is signaled by a vanishing of $B_M$ (see panels $B$ to $D$). At even lower densities, when $n\lesssim1.70$, the sample turns highly insulating and exhibits non-linear $I$-$V$, suggestive of a pinned Wigner solid. (\textit{B}) Our experimental technique to measure spin polarization of the 2DES, demonstrated using data taken at $n=2.63$ and density-of-states diagrams. The sample resistance $R_{[100]}$, measured along the [100] direction (see inset), increases as the magnetic field $B_{||}$ applied parallel to the 2DES plane polarizes the electron spins, and then saturates once the system is fully magnetized at the field $B_M$. Above $B_M$, the Zeeman energy ($E_Z$) exceeds the Fermi energy ($E_F$); $g(E)$ is the density of states. (\textit{C}) Resistance ($R_{[100]}$) vs. $B_{||}$ data taken at different 2DES densities ($R_{[010]}$ data show a similar behavior). Traces are shown for both up- and down-sweeps of $B_{||}$ and are offset vertically. The vertical arrows mark the positions of $B_M$ above which the resistance saturates. They are placed symmetrically with respect to $B_{||} = 0$, and are based on the average of four values of $B_M$ (for up- and down-sweeps, and $+B_{||}$ and $-B_{||}$). For $n \leq 2.00$, the traces are flat and there is no hint of a $B_M$. (\textit{D}) $B_M$ and (\textit{E}) the deduced spin susceptibility are plotted as a function of density (lower axis) and $r_s$ (top axis). All measurements were performed at $T=0.30$ K.
}
\end{figure*}

\begin{figure*}[t!]
\includegraphics[width=1\textwidth]{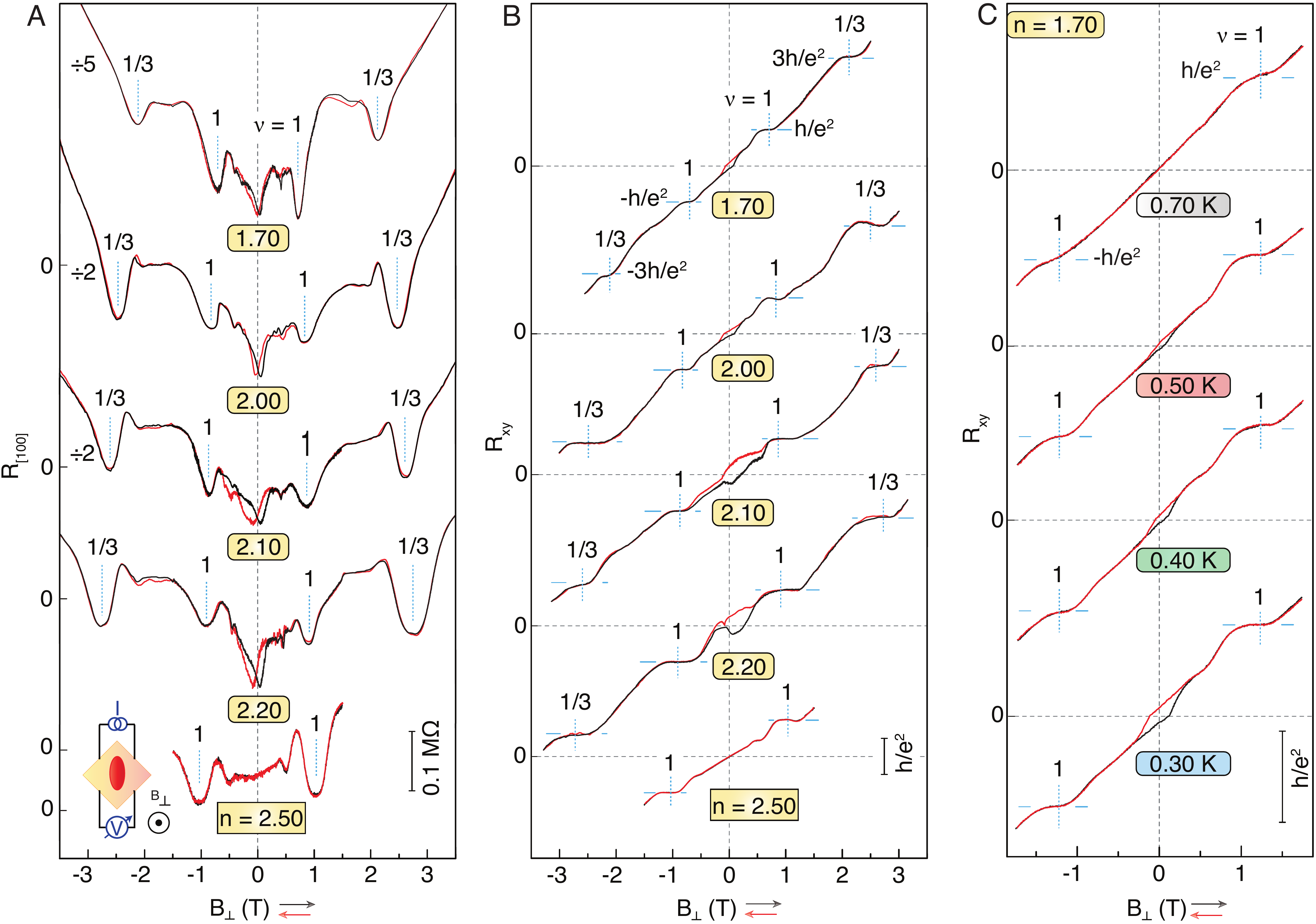}
\caption{\label{fig:fig2} Magnetotransport data near $n=n_c$. (\textit{A}) Longitudinal ($R_{[100]}$) and (\textit{B}) Hall ($R_{xy}$) resistances, measured at $T=0.30$ K, are shown as a function of \textit{perpendicular} magnetic field ($B_{\perp}$) for five densities as indicted (in units of $10^{10}$ cm$^{-2}$). As the density is lowered, a pronounced hysteresis near $B_{\perp}=0$ emerges close to $n_c = 2.00$ and continues at $n<n_c$. The longitudinal resistance traces show deep minima at Landau level filling factors $\nu=1$ and $1/3$, accompanied by nearly quantized plateaus in $R_{xy}$ at the expected values, signaling  well-developed quantum Hall states at these fillings. The presence of fractional quantum Hall states at densities as low as $n=1.70$ demonstrate the extremely high quality of the 2DES, and the presence of electron-electron interaction. (\textit{C}) Temperature dependence of $R_{xy}$, taken at $n=1.70$, shows that the hysteresis vanishes at high temperatures ($T \simeq 0.70$ K). 
 } 
\end{figure*}

\begin{figure*}[t!]
\includegraphics[width=1\textwidth]{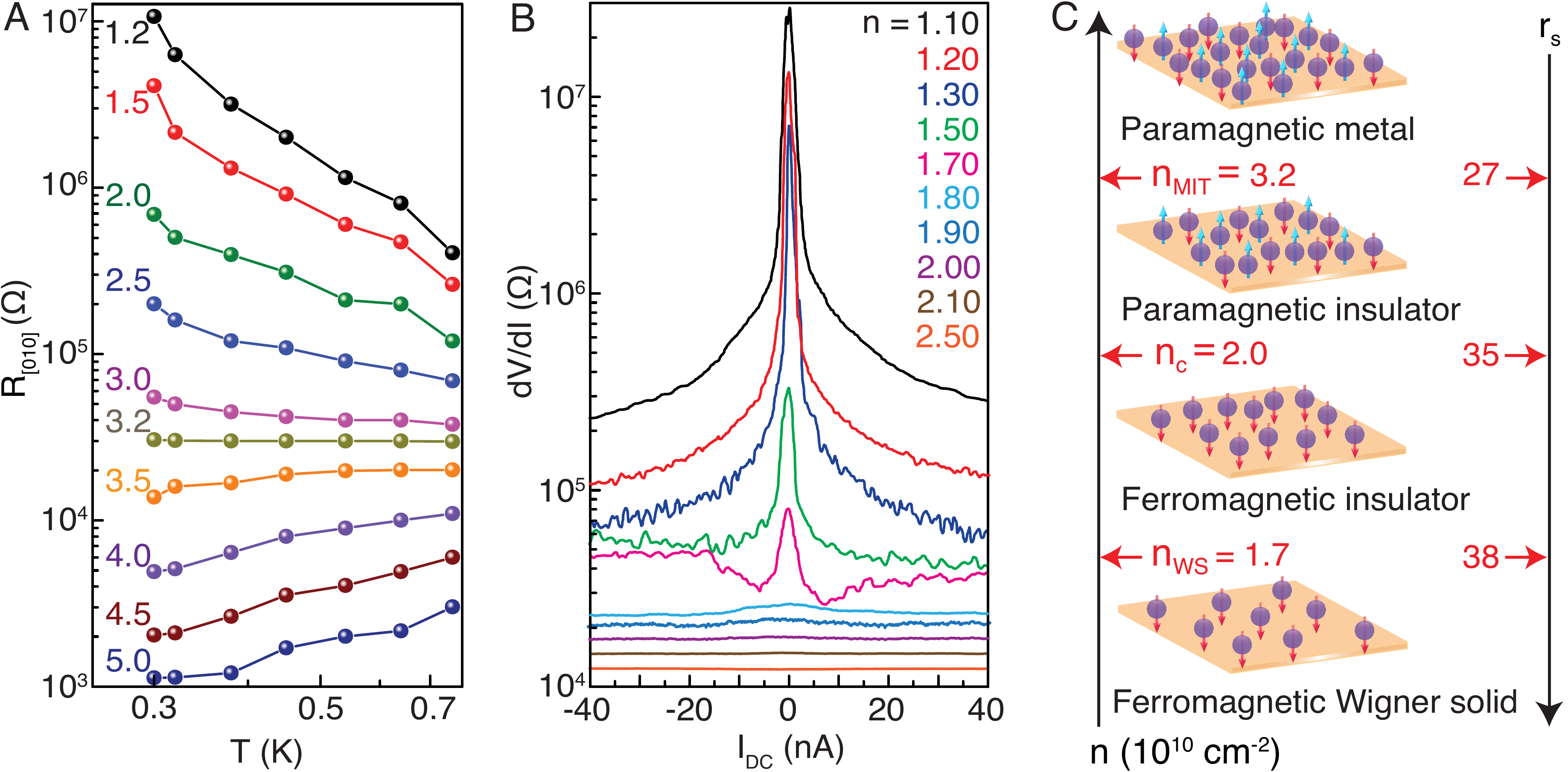}
\caption{\label{fig:fig3} Summary of temperature-dependent and non-linear $I$-$V$ data, and the phases that we observe as a function of $n$ (or $r_s$). (\textit{A}) $R_{[010]}$, resistance measured along [010], as a function of temperature exhibiting metallic transport for $n \geq 3.2$, and a switch over to an insulating behavior for smaller $n$; $R_{[100]}$ data show a qualitatively similar behavior. (\textit{B}) Differential resistance ($dV/dI$) measured along [100] at $T=0.30$ K, plotted against the DC bias current, showing increasing non-linearity at small biases when $n\lesssim 1.70$. (\textit{C}) Illustration of different phases observed in our 2DES. Starting as a paramagnetic metal at large $n$ (small $r_s$), the 2DES undergoes a cascade of transitions to a paramagnetic insulator, ferromagnetic insulator, and finally to a ferromagnetic solid as the density is lowered.} 
\end{figure*}

\begin{figure*}[t!]
\includegraphics[width=0.99\textwidth]{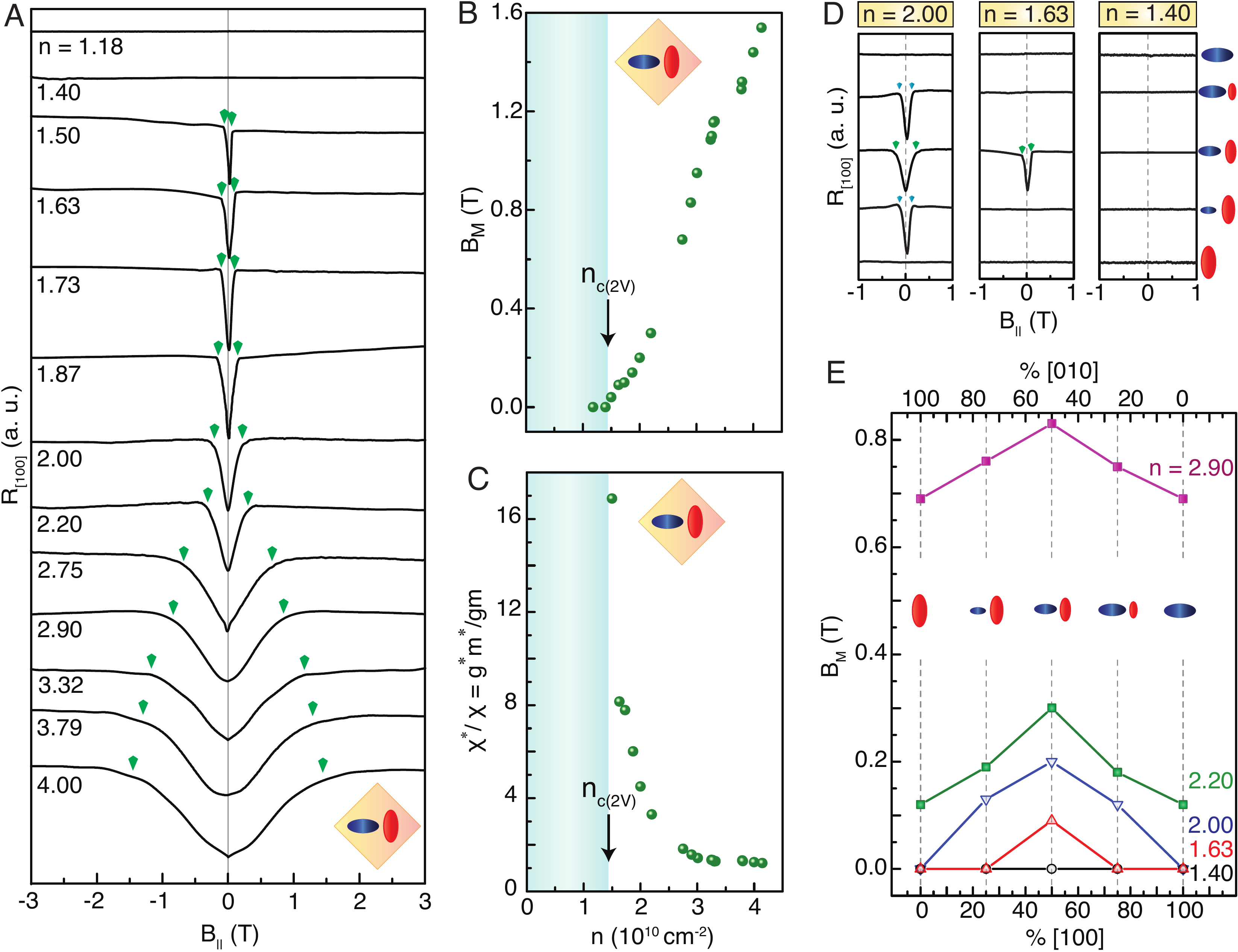}
\caption{\label{fig:fig4} Controlling the spin polarization and ferromagnetic transition by tuning the valley occupancy. (\textit{A}) Resistance vs. $B_{||}$ data taken at $T = 0.30$ K for different 2DES densities when the two occupied valleys are degenerate. The vertical arrows mark the positions of $B_M$. They are placed symmetrically with respect to $B_{||}$, and are based on the average of two values of $B_M$ (for $+B_{||}$ and $-B_{||}$). For $n \leq 1.40$, the traces are flat and there is no hint of a $B_M$. (\textit{B}), (\textit{C}) $B_M$ and the deduced spin susceptibility of the valley-degenerate 2DES are plotted as a function of density. (\textit{D}) Resistance vs. $B_{||}$ data taken at three 2DES densities demonstrating the effect of the valley occupancy on $B_M$ (marked with vertical arrows). The valley occupancy is graphically indicated on the right. At a fixed density, the 2DES has the smallest $B_M$ when only one valley (either [100] or [010]) is occupied. As electrons transfer from one valley to the other, $B_M$ increases and reaches its maximum value when the valleys are equally occupied. The single-valley case reaches the ferromagnetic transition at the largest density while the valley-degenerate case requires the smallest density. (\textit{E}) $B_M$ plotted against the valley occupancy, showing a clear increase in $B_M$ as we move from single-valley cases ($0\%$ and $100\%$) towards the valley-degenerate ($50\%$) case.}
\end{figure*}

There are other semiconductor 2DESs with large $m$, e.g., at a Si/SiO$_2$ or MgZnO/ZnO interface, or in AlAs quantum wells, where large $r_s$ values can be reached  \cite{Okamoto.PRL.1999, Vitkalov.PRL.2001, Shashkin.PRL.2001, Vakili.PRL.2004, Gunawan.nphys.2007, Kravchenko.Rep.Prog.Phys.2004, Spivak.Rev.Mod.Phys.2010, Pudalov.PRB.2018, Li.PRB.2019, Falson.RPP.2018}. Indeed, the spin polarization of interacting 2DESs became a subject of intense interest and controversy in the context of the enigmatic metal-insulator transition (MIT) in dilute 2D carrier systems. Numerous experiments revealed that the spin/valley polarization in 2D carrier systems plays a role in the temperature dependence of conductivity \cite{Papadakis.Science.1999, Tutuc.PRL.2001, Okamoto.PRL.1999, Vitkalov.PRL.2001, Shashkin.PRL.2001, Zhu.PRL.2003, Winkler.PRB.2005, Vakili.PRL.2004, Gunawan.nphys.2007, Kravchenko.Rep.Prog.Phys.2004, Spivak.Rev.Mod.Phys.2010, Pudalov.PRB.2018, Li.PRB.2019}. In Si/SiO$_2$ 2DESs, there were also claims that the spin susceptibility diverges at $r_s\simeq 9$, and that the divergence coincides with the MIT \cite{Vitkalov.PRL.2001, Shashkin.PRL.2001}. However, these conclusions were not corroborated in measurements on a nearly-ideal (single-valley, isotropic, very thin) 2DES confined to a narrow (4.5-nm-wide) AlAs quantum well \cite{Vakili.PRL.2004}. The AlAs data showed that the spin susceptibility does not diverge up to the highest experimentally achieved $r_s$ ($\simeq10$). Instead, it closely follows the Monte Carlo calculations' results \cite{Attaccalite.PRL.2002} and remains finite as the 2DES goes through the MIT at $r_s \simeq 8$ \cite{Vakili.PRL.2004}. In very recent studies which revisited the MIT problem in Si/SiO$_2$ 2DESs \cite{Pudalov.PRB.2018, Li.PRB.2019}, it was also concluded that there is no divergence of the spin susceptibility, even at densities lower than those reached in Refs. \cite{Vitkalov.PRL.2001, Shashkin.PRL.2001}. Therefore, there has been no experimental evidence for a transition to full spin polarization in a 2DES at zero magnetic field prior to this work.

\section*{Experimental Results}
We report here the observation of a spontaneous transition to a fully-magnetized ground state and experimentally construct a comprehensive ground-state phase diagram for the interacting 2DES as a function of electron density (and $r_s$), as summarized in Fig. \ref{fig:fig1}\textit{A}. Our material platform is a very low disorder, dilute 2DES confined to a modulation-doped, 21-nm-wide AlAs quantum well ($\varepsilon=10$) \cite{Chung.PRM.2018}. The 2D electrons in our sample occupy two in-plane conduction-band valleys with longitudinal and transverse effective masses $m_l = 1.1$, and $m_t=0.20$, leading to an effective in-plane band mass $m$ equal to $(m_l m_t)^{1/2}=0.46$ \cite{Shayegan.AlAs.Review.2006}. We refer to these valleys, whose electron occupancy we can control via the application of in-plane, uniaxial strain, by the direction of their major axis in the plane, [100] and [010] (see Section I of \textit{SI Appendix}). The large effective mass, combined with the exceptionally high purity of the samples, allows us to achieve very large values of $r_s$ while maintaining a strongly interacting system. This is evinced by our observation of signatures of fractional quantum Hall states in a perpendicular magnetic field a at density as low as $n=1.20$ ($r_s \simeq 45$) (see, e.g. Fig. \ref{fig:fig2} for data at $n=1.70$, and relevant discussion); throughout the manuscript, we give the 2DES densities in units of $10^{10}$ cm$^{-2}$.

\subsection*{Signatures of Spontaneous Ferromagnetism}
First we present data for the case where all the electrons are transferred to the [010] valley, i.e., the valley whose longer Fermi wavevector axis is along the [010] direction; see Fig. \ref{fig:fig1}\textit{B} lower-panel inset (for more details, see Section I of \textit{SI Appendix}). We probe the spin polarization via measuring the 2DES resistance as a function of a magnetic field $B_{||}$ applied \textit{parallel} to the sample plane (Fig. \ref{fig:fig1}\textit{B}). This is a well-established technique \cite{Okamoto.PRL.1999, Vitkalov.PRL.2001, Shashkin.PRL.2001, Zhu.PRL.2003, Tutuc.PRL.2001, Winkler.PRB.2005, Vakili.PRL.2004, Gunawan.nphys.2007, Kravchenko.Rep.Prog.Phys.2004, Spivak.Rev.Mod.Phys.2010, Pudalov.PRB.2018, Li.PRB.2019, Dolgopolov.2000, Herbut.2001} whose working principle, as depicted in Fig. \ref{fig:fig1}\textit{B}, is that the resistance of the 2DES increases as $B_{||}$ polarizes the electrons through the addition of Zeeman energy ($E_Z$). The partially-polarized 2DES has higher resistance because of a reduction in the screening of the disorder potential by the 2D electrons \cite{Dolgopolov.2000, Herbut.2001}. When $B_{||}$ reaches a sufficiently large value ($B_M$) so that $E_Z$ equals the Fermi energy ($E_F$), the 2D electrons become fully magnetized and the resistance no longer rises. Since $E_Z=g \mu_B B$, and $E_F =(2 \pi \hbar^2 n)/m$, we have:
\begin{equation} \label{eq:1}
 B_M = \bigg(\frac{2\pi \hbar^2}{\mu_B}\bigg)\frac{n}{gm}, 
\end{equation} 
where $g$ is the band value of the effective Land\'e \textit{g}-factor and $\mu_B$ is the Bohr magneton ($g=2$ in our AlAs 2DES). Note that $B_M$ is inversely proportional to the product $gm$ which is directly related to the spin susceptibility, $\chi$. In an interacting 2DES, $g$, $m$, and $\chi$ are typically enhanced by interaction. Denoting the enhanced values by $g^*$, $m^*$, and $\chi^*$, one can determine the enhancement factor for the susceptibility $\chi^*/\chi=g^*m^*/gm$ from measuring $B_M$. In our experiments, we find that as we lower the 2DES density, $B_M$ decreases rapidly, signaling a fast-enhancing $\chi^*/\chi$, and then \textit{suddenly vanishes} below a critical density $n_c$ (Figs. \ref{fig:fig1}\textit{C}-\textit{E}). This observation strongly suggests that the 2DES is fully magnetized for $n \leq n_c$ in the absence of an applied magnetic field.

Figures \ref{fig:fig1}\textit{C}-\textit{E} capture the evolution described above. In Fig. \ref{fig:fig1}\textit{C} we show $R_{[100]}$ vs. $B_{||}$ applied along the [100] direction (see Fig. \ref{fig:fig1}\textit{B} inset). For $n \geq 2.10$, $R_{[100]}$ rises with $B_{||}$ and saturates above a density-dependent field $B_M$, marked by vertical arrows in Fig. \ref{fig:fig1}\textit{C}. As discussed above (Fig. \ref{fig:fig1}\textit{B}), $B_M$ signals the full magnetization of the 2DES and allows us to deduce the spin susceptibility. In Figs. \ref{fig:fig1}\textit{D}, \textit{E} we show plots of the measured $B_M$ and the deduced susceptibility as a function of $n$. As expected for a dilute, interacting 2DES, when $n$ is lowered, $B_M$ decreases quickly as susceptibility increases \cite{Attaccalite.PRL.2002, Vakili.PRL.2004}. The principal novel finding in our experiments is that, as $n$ is lowered below $2.10$, the traces in Fig. \ref{fig:fig1}\textit{C} become completely independent of $B_{||}$ with no sign of a $B_M$ whatsoever. A most logical interpretation of this behavior is that, for $n \leq 2.00$, the 2DES is fully magnetized at $B_{||}=0$, expected for spontaneous ferromagnetism. The critical density $n_c=2.00$ corresponds to $r_s=34.5$. 

It is noteworthy in Fig. \ref{fig:fig1}\textit{C} that the trace taken at a slightly larger density, $n=2.10$, exhibits a pronounced hysteresis near $B_{||}=0$. In Figs. \ref{fig:fig2}\textit{A}, \textit{B} we show $R_{[100]}$ and $R_{xy}$ (Hall resistance) traces taken as a function of $perpendicular$ magnetic field, $B_{\perp}$. Hysteretic features near zero magnetic field are clearly visible here too, and are most pronounced when $n$ is close to $n_c$. These features suggest a possible formation of magnetic domains, as expected near a magnetic transition. Notably, as seen in Fig. \ref{fig:fig2}\textit{C}, the hysteresis vanishes at high temperatures; e.g. at $n=1.70$, the hysteresis disappears at $T \simeq 0.70$ K.

We note that the technique we use to probe the enhancement and divergence of the spin susceptibility is similar to what was used in many previous studies of the spin-polarization of dilute 2D carriers \cite{Okamoto.PRL.1999, Vitkalov.PRL.2001, Shashkin.PRL.2001, Zhu.PRL.2003, Tutuc.PRL.2001, Winkler.PRB.2005, Vakili.PRL.2004, Gunawan.nphys.2007, Kravchenko.Rep.Prog.Phys.2004, Spivak.Rev.Mod.Phys.2010, Pudalov.PRB.2018, Li.PRB.2019, Dolgopolov.2000, Herbut.2001}. None of the previous studies, however, reported a sudden and complete disappearance of the magnetoresistance as a function of $B_{||}$ (implying a vanishing $B_M$ and full magnetization) as the 2DES density is reduced below a critical value. In some of these studies, e.g., in the work of Ref. \cite{Shashkin.PRL.2001} on Si$/$SiO$_{2}$, non-zero magnetoresistance was observed down to the lowest achievable densities, and a plot of $B_M$ (deduced from fitting the magnetoresistance to an empirical expression) vs. $n$ was \textit{extrapolated} to lower densities to conclude that $B_M$ vanishes at a finite density (corresponding to an $r_s \simeq 9$). As shown in Ref. \cite{Vakili.PRL.2004}, such extrapolation can be misleading, and recent new measurements \cite{Pudalov.PRB.2018, Li.PRB.2019} indeed conclude that there is no magnetic instability or complete spin polarization at zero magnetic field in the Si$/$SiO$_{2}$ system even at densities lower than those achieved in Ref. \cite{Shashkin.PRL.2001} [see, e.g., point 7 in page 10 of Ref. \cite{Pudalov.PRB.2018}].

We believe it is the exceptionally high sample quality that allows us to explore the interaction phenomena even in the extremely dilute case and observe a spontaneous magnetization. The high quality of our 2DES can be inferred from Fig. \ref{fig:fig2}\textit{A}, \textit{B} traces. At Landau level filling factor $\nu=1$ and $1/3$ ($\nu=nh/eB_{\perp}$), we observe clear indications of integer and fractional quantum Hall states, signaled by resistance minima in $R_{[100]}$ minima and reasonably well-quantized plateaus in $R_{xy}$ even at $n=1.70$ ($r_s \simeq 38$). [We observe $R_{[100]}$ minima at $\nu=1$ and $1/3$ down to $n=1.20$ ($r_s \simeq 45$).] The observation of fractional quantum Hall states down to such small densities is particularly noteworthy as it attests to the presence of interaction even when the 2DES is extremely dilute.

We would like to highlight two points before closing this section. First, note that Eq. \ref{eq:1} assumes that $gm$ (i.e., the spin susceptibility, $\chi$) is a constant, independent of $B_{||}$. This assumption is consistent with the conclusions of previous studies \cite{Vakili.PRL.2004}. While our determination of $\chi^*$ relies on this assumption, we emphasize that our main conclusion, namely that the 2DES undergoes a spontaneous spin polarization below a critical density, is based on the vanishing of $B_{M}$, and does not depend on such an assumption. Second, as shown in Section IV of the \textit{SI Appendix}, the value of $R_{[100]}$ (which is constant and independent of $B_{||}$) at $n=2.00$ is close to the saturated value of $R_{[100]}$ at \textit{large $B_{||}$ at $n=2.10$}. This observation confirms our conclusion that the 2DES is fully spin polarized at $n=2.00$ at $B_{||}=0$, just as it is at $n=2.10$ at large $B_{||}$.

\subsection*{Is There a Link Between the Spontaneous Ferromagnetism \& the Metal-Insulator Transition?}
As mentioned in the opening paragraphs, a possible link between the spin polarization and MIT in dilute 2D carrier systems has been debated heavily \cite{Papadakis.Science.1999, Tutuc.PRL.2001, Okamoto.PRL.1999, Vitkalov.PRL.2001, Shashkin.PRL.2001, Zhu.PRL.2003, Winkler.PRB.2005, Vakili.PRL.2004, Gunawan.nphys.2007, Kravchenko.Rep.Prog.Phys.2004, Spivak.Rev.Mod.Phys.2010, Pudalov.PRB.2018, Li.PRB.2019}. In Fig. \ref{fig:fig3}\textit{A} we show the temperature dependence of the sample resistance measured at $B=0$. At the highest $n$, the behavior is ``metallic," with resistance decreasing with decreasing temperature, while at the lowest densities, an opposite, ``insulating," behavior is seen. The density at which the behavior switches is $n_{MIT} \simeq 3.2$ ($r_s \simeq 27$), well above $n_c=2.0$, the onset of the transition to full magnetization. This is again consistent with the data of Ref. \cite{Vakili.PRL.2004} but disagrees with the conclusions of Ref. \cite{Shashkin.PRL.2001}. It is also noteworthy that $r_s \simeq 27$ at which we observe the MIT is much larger than $r_s$ reported in other 2D carrier systems for the MIT, except for very clean GaAs 2D hole systems \cite{Yoon.PRL.1999}.

We would like to emphasize that, while we do not see a link between the MIT and the full magnetization of the 2DES (as inferred from the sudden disappearance of $B_{M}$ in Fig. \ref{fig:fig1}\textit{C} data), we do observe certain anomalies in the magnetoresistance traces near $n_{MIT} \simeq 3.2$. As seen in Fig. \ref{fig:fig1}\textit{C}, at $n=3.30$, just above the MIT, the resistance shows a very rapid rise as a function of $B_{||}$ at very low fields, followed by an unusual \textit{linear} rise with $B_{||}$ at higher fields before saturation. This behavior is not unique to this trace, and can be seen in a range of densities between $\simeq 3.60$ and $\simeq 3.05$, near the MIT (see Fig. S4 of \textit{SI Appendix}). Another noteworthy feature is a change in the slope of $B_M$ vs. $n$ plot in Fig. \ref{fig:fig1}\textit{D} at $n_{MIT}$: it appears that the $B_M$ drop with lowering the density accelerates below $n_{MIT}$. A qualitatively similar feature was reported recently for Si/SiO$_2$ 2DESs  \cite{Li.PRB.2019}. We will return to these anomalous features later in the manuscript in our interpretation of the data.

\subsection*{Hints of Wigner Solid}
In Fig. \ref{fig:fig3}\textit{B}, we address another fundamental property of our 2DES at very low densities. Figure \ref{fig:fig3}\textit{B} displays differential resistance ($dV/dI$) as a function of applied DC current $I_{DC}$; see Sections V-VII of the \textit{SI Appendix} for measurement details and additional data. At densities $n \geq n_c$, $dV/dI$ is linear. As the density is lowered below $n_c$, $I$-$V$ slowly becomes non-linear and, at the lowest densities, it is strongly non-linear. As demonstrated in Fig. S11 of the \textit{SI Appendix}, this non-linearity is a strong function of temperature, and vanishes at high temperatures. Previous studies on extremely dilute 2D carrier systems at $B=0$ \cite{Yoon.PRL.1999}, or at very small Landau level fillings \cite{Goldman.PRL.1990}, have concluded that the non-linear $I$-$V$ suggests the depinning of a Wigner solid which is pinned by the ubiquitous disorder potential. This conclusion has been corroborated by numerous experimental studies, including microwave resonance and other measurements \cite{Williams.PRL.1991, Chen.Nature.Phys.2006, Tiemann.Nat.Phys.2014, Deng.PRL.2016}. It is possible that in our 2DES, too, the non-linear $I$-$V$ signals the formation of a pinned Wigner solid at $r_s  \gtrsim 38$ ($n \lesssim 1.70$). 

An intriguing observation in our experiments is that the fractional quantum Hall states still show up in a density regime where the longitudinal resistance at $B=0$ is extremely large $>> h/e^2$ and the 2DES exhibits a strongly insulating temperature dependence. The resistance is also extremely large at $\nu=1/3$ but, remarkably, its value at $\nu=1/3$ \textit{decreases} as we lower the temperature, as expected for a fractional quantum Hall liquid state. The notion that a pinned Wigner solid at $B=0$ would make a transition to a correlated liquid state in a perpendicular field might sound counterintuitive, but in fact is consistent with conclusions implied by the theoretical work of Zhao \textit{et al.} \cite{Zhao.PRL.2018}.

\subsection*{Phase Diagram for an Interacting 2D Electron System}
The observations of metal-insulator transition, spontaneous magnetization, and non-linear $I$-$V$ response as a function of $n$ (or $r_s$) lead us to suggest an experimental phase diagram for the ground states of our interacting 2DES, as highlighted in Fig. \ref{fig:fig1}\textit{A}. The ground states include four distinct phases as a function of decreasing $n$ (i.e. increasing $r_s$), as schematically represented in Fig. \ref{fig:fig3}\textit{C}: \textit{(i)} paramagnetic metal, \textit{(ii)} paramagnetic insulator, \textit{(iii)} ferromagnetic insulator, and \textit{(iv)} ferromagnetic Wigner solid.

\subsection*{Interpretation of the Ferromagnetic Transition}
A simple but possibly na\"ive interpretation of the observed ferromagnetic transition we observe is that it is a Bloch transition \cite{Bloch.1929}. As alluded to in the introduction, in 1929 Felix Bloch suggested that itinerant electrons should spontaneously magnetize at sufficiently low densities. Our observation of the spontaneous ferromagnetism is highly reminiscent of such phenomenon. A transition at lower densities to a state that we identify as a Wigner solid would also make qualitative sense in this simple picture. Such a picture is also qualitatively consistent with the Monte Carlo calculations of \cite{Attaccalite.PRL.2002} which were done for a disorder-free 2DES, although there are quantitative discrepancies. For example, the value of $r_s$ ($\simeq 35$) at which we observe the ferromagnetic transition is larger than the value ($\simeq 26$) predicted by Ref. \cite{Attaccalite.PRL.2002}. So is the value of $r_s$ ($\simeq 38$) above which we see signs of a possible Wigner solid formation (Ref. \cite{Attaccalite.PRL.2002} predicts $r_s\simeq 35$). These quantitative discrepancies might come from the fact that the calculations are performed for an \textit{ideal} 2DES, with no disorder, zero electron layer thickness, and an isotropic effective mass. In our 2DES, on the other hand, the electrons have a non-zero layer thickness and occupy a valley with an anisotropic effective mass. Both the finite electron layer thickness and the mass anisotropy are known to reduce the interaction strength and the enhancement of the spin susceptibility at a given $r_s$ \cite{Gokmen.PRB.2007}, and can move the onset of full magnetization to larger $r_s$.

However, there is an important caveat. As clean as our sample might be, there is some finite disorder because of the presence of ionized impurities. The disorder is indeed the likely cause for the 2DES exhibiting an ``insulating" behavior below $n_{MIT}$. On the other hand, our observation of a spontaneous ferromagnetism, as well as the fractional quantum Hall features that we observe at $n_c$ and even lower densities (Fig. \ref{fig:fig2}) provide strong evidence that interaction \textit{is} playing a dominant role. An alternative interpretation of our data is then as follows. It has been shown theoretically that a direct first-order transition between a metallic Fermi liquid state and a Wigner solid is forbidden in 2D, and that there should be one or more exotic ``mirco-emulsion" phases in an intermediate density range \cite{Spivak.Rev.Mod.Phys.2010, Spivak.2004}. It might be that the ferromagnetic transition we observe is occurring in such a micro-emulsion phase. The anomalous features seen in the resistance vs. $B_{||}$ data near the MIT (Fig. \ref{fig:fig1}\textit{C} and Fig. S4 of \textit{SI Appendix}), could be manifestations of such exotic intermediate phases. Also, given the anisotropy of the effective mass in our system and that the resistance we measure in this phase is anisotropic ($R_{[100]} < R_{[010]}$), it is possible that the intermediate phase is nematic or smectic as proposed theoretically \cite{Spivak.Rev.Mod.Phys.2010, Spivak.2004}. Moreover, the theory of Refs. \cite{Spivak.2004, Chakravarty.1999} concludes that the Wigner solid phase stabilized at the lowest densities has ferromagnetic order. This scenario is consistent with our data: we see $B_M=0$ even below $n \simeq 1.70$ where we infer a transition to a pinned Wigner solid phase in our 2DES.

\subsection*{Role of the Valley Degree of Freedom in the Spin Polarization}
Another important highlight of experiments is captured in Fig. \ref{fig:fig4} where we illustrate how the valley polarization impacts the ferromagnetic transition. How we tune the valley occupancy is described in Section I of the \textit{SI Appendix}. In Fig. \ref{fig:fig4}\textit{A} we show a plot similar to Fig. \ref{fig:fig1}\textit{C} except that here the valleys are degenerate. We also show plots of the measured $B_M$ and the deduced spin susceptibility as a function of $n$ in Figs. \ref{fig:fig4}\textit{B}, \textit{C}. Qualitatively similar to Figs. \ref{fig:fig1}\textit{C}-\textit{E}, when $n$ is lowered, $B_M$ decreases rapidly as susceptibility increases. However, in the valley-degenerate case, the critical density below which $B_M$ vanishes is $n_{c(2V)} = 1.40$ instead of $n_{c(1V)} = 2.00$ of the single-valley-occupied 2DES (Figs. \ref{fig:fig1}\textit{C}-\textit{E}). This finding is consistent with previous measurements (at much higher densities) \cite{Shkolnikov.PRL.2004, Gokmen.PRB.2010} and theoretical calculations \cite{Marchi.PRB.2009} which concluded that the enhancement of the spin susceptibility, at a given density, is smaller for a two-valley system compared to a single-valley one. The phenomenon can be attributed to the modification of the exchange energy in the system because of the extra (valley) degree of freedom. 

The tunability of the valley occupancy allows us to investigate the ferromagnetic transition as a function of partial valley occupancy. As shown in Figs. \ref{fig:fig4}\textit{D},\textit{E}, $B_M$ is largest when the valleys are degenerate. As we lift the degeneracy and move towards valley polarization, $B_M$ starts to drop and eventually attains its smallest value when the electrons are completely valley polarized. This means that the spin susceptibility is maximum when the 2DES is valley-polarized and minimum when the valleys are degenerate, for a given density. Thanks to this interplay between the spin and valley degrees of freedom, we can control the onset of the ferromagnetic transition via tuning the valley occupancy, opening up exciting avenues to integrate spintronics and valleytronics in the same device.

\begin{acknowledgments}
We acknowledge support through the U.S. Department of Energy Basic Energy Science (Grant No. DEFG02-00-ER45841) for measurements, and the National Science Foundation (Grants No. DMR 1709076), No. ECCS 1906253, and No. MRSEC DMR 1420541), and the Gordon and Betty Moore Foundation’s EPiQS Initiative (Grant No. GBMF9615 to L. N. P.) for sample fabrication and characterization. We also thank D. M. Ceperley, H. D. Drew, J. K. Jain, S. A. Kivelson, M. A. Mueed, and M. P. Sarachik for illuminating discussions.

\end{acknowledgments}

\end{document}